\documentclass[times,authoryear]{elsarticle}
\usepackage{jasr}
\usepackage{framed,multirow}

\usepackage{amssymb}
\usepackage{latexsym}
\usepackage{amsmath}
\usepackage{graphicx}

\usepackage{url}
\usepackage{xcolor}
\definecolor{newcolor}{rgb}{.8,.349,.1}

\usepackage[citebordercolor=white]{hyperref}

\journal{Advances in Space Research}

\begin{document}

\verso{Z. Liu \textit{et al.}}

\begin{frontmatter}

\title{FIVR-NLFFF: A fully-implicit viscous-relaxation code for nonlinear force-free magnetic field extrapolation of the solar corona}%
%\tnotetext[tnote1]{This is an example for title footnote coding.}

\author{Zhenhua \snm{Liu}}
\author{Chaowei \snm{Jiang}\corref{cor1}}
%\fntext[fn1]{This is author footnote for second author.}
% \author[2]{Given-name3 \snm{Surname3}}
% %% Third author's email
\cortext[cor1]{Corresponding author.}
\ead{chaowei@hit.edu.cn}
% \author[2]{Given-name4 \snm{Surname4}}

%\address[2]{Affiliation 1, Address, City and Postal Code, Country}
\affiliation{organization={State Key Laboratory of Solar Activity and Space Weather, School of Aerospace Science, Harbin Institute of Technology, Shenzhen},
                addressline={Nanshan},
                city={Shenzhen},
                postcode={518000},
                country={China}}

%\address[2]{Affiliation 2, Address, City and Postal Code, Country}
% \affiliation[2]{organization={Affiliation 2},
%                 addressline={Address},
%                 city={City},
%                 postcode={Postal Code},
%                 country={Country}}

% \received{1 May 2013}
% \finalform{10 May 2013}
% \accepted{13 May 2013}
% \availableonline{15 May 2013}
% \communicated{S. Sarkar}

\begin{abstract}
Magnetic field extrapolation from the solar photosphere to the corona plays an important role in solar physics research. In this work, we present a fully-implicit viscous-relaxation nonlinear force-free field (FIVR-NLFFF) extrapolation code based on a viscous magnetohydrodynamic relaxation model. The method solves the magnetic induction equation alongside a simplified momentum equation, which assumes a balance between the Lorentz force and the viscous force. Under this assumption, the velocity field driving the magnetic field evolution is determined instantaneously by the Lorentz force distribution. Through viscous dissipation, the system relaxes toward a minimum-energy state, consistent with the vector magnetogram prescribed at the lower boundary. To enhance numerical stability, we adopt a fully implicit time integration scheme and employ central finite differences for spatial discretization. The resulting system of nonlinear algebraic equations is solved using the Jacobian-free Newton-Krylov method, as implemented in the Portable, Extensible Toolkit for Scientific Computation (PETSc). We validate the code using three benchmark models: the Low and Lou force-free solution, the Titov-Démoulin magnetic flux rope model, and a strongly sheared arcade configuration containing a current sheet. Quantitative comparisons demonstrate good agreement with the reference solutions. Notably, the code's ability to handle discontinuities and reconstruct coronal current sheets makes it a promising tool for studying magnetic fields that may directly trigger solar eruptions.
\end{abstract}

\begin{keyword}
%% MSC codes here, in the form: \MSC code \sep code
%% or \MSC[2008] code \sep code (2000 is the default)
%\MSC 41A05\sep 41A10\sep 65D05\sep 65D17
%% Keywords
\KWD Magnetic fields\sep Sun: corona\sep Methods: numerical
\end{keyword}

\end{frontmatter}

%% For linenumbers
% \linenumbers

%% main text
\section{Introduction}
\label{sec:Introduction}
It is widely acknowledged that magnetic fields play the dominant role in the structure and evolution of solar corona. In particular, the coronal magnetic field drives solar eruptive phenomena, including flares and coronal mass ejections \citep{priest2002magnetic}. Therefore, understanding the three-dimensional (3D) configuration of the coronal magnetic field is crucial for studying the mechanisms of these activities. However, it remains challenging to measure the coronal magnetic field such that the 3D magnetic configurations can be fully recovered, although important progresses have been made in the recent years \citep{yang2020global,yang2024observing}. A common practice in research is to extrapolate the coronal magnetic field from the measurable photospheric magnetic field (which is known as solar magnetograms). One of the most frequently used models is the Lorentz force-free field model, which is described as:
\begin{align} 
    \mathbf{J} \times \mathbf{B} = 0, \quad \nabla \cdot \mathbf{B} = 0,
\end{align}
where $\mathbf{J}= \nabla \times \mathbf{B}$ is the current density and $\mathbf{B}$ the magnetic field. The underlying justification for the force-free model is that, since the magnetic field pressure gradient is much greater than the plasma thermal pressure gradient and other non-magnetic forces in the low corona, the magnetic pressure gradient must be balanced by the magnetic tension force, and therefore the Lorentz force (which is the sum of magnetic pressure gradient force and tension force) disappears. Note that this assumption is only reasonable in the static corona before solar eruption initiation, and furthermore, the force-free assumption is generally not valid in the photosphere and the low atmosphere where the plasma $\beta$ (the ratio of gas to magnetic pressure) is often larger than unity. Based on the assumptions for distribution of current in space, force-free fields can be classified into potential field (which is current free), linear force-free field (which assumes constant ratio between current and magnetic field), and nonlinear force-free field (NLFFF). The first two types can be straightforwardly obtained by solving the Poisson equation and the Helmholtz equation \citep{jiang2012unified}, respectively. As a more accurate model for reconstructing the coronal magnetic field, the general equation of NLFFF is much more difficult to solve due to its nonlinear nature.

In the past decades, a variety of numerical methods have been proposed to solve the NLFFF with photospheric vector magnetograms as the bottom boundary condition. As reviewed by \citet{wiegelmann2021solar}, the numerical methods are classified into the five categories: the upward integration method \citep{nakagawa1974dynamics, wu1990numerical}, the Grad-Rubin method \citep{grad1958hydromagnetic, sakurai1981calculation}, the magnetohydrodynamic (MHD) relaxation method \citep{chodura19813d, mikic1988dynamical,jiang2011reconstruction,inoue2013nonlinear,hayashi2018mhd,price2019time,lumme2022data}, the optimization approach \citep{wheatland2000optimization,wiegelmann2004optimization,wiegelmann2007computing}, and the boundary-element method \citep{yan2000new}. The Grad-Rubin, MHD relaxation, and optimization methods are of iterative type, in which often a potential field is firstly computed, and then the non-potential field part is incrementally added up through a number of iteration steps such that both the bottom boundary requirement and the force-free constraint are met by the final solution. Although a potential field is typically used as the initial condition, other models can also be considered, and the differences in NLFFF solutions obtained from different initial guesses have been discussed \citep{hayashi2022nonpotentiality, barnes2024electric}. In contrast, upward integration and boundary-element methods directly solve the NLFFF equations without first computing a potential field. \citet{schrijver2006nonlinear} and \citet{metcalf2008nonlinear} compared a number of different NLFFF extrapolation codes using a semi-analytic force-free solutions proposed by \citet{low1990modeling} and a solar magnetic flux rope model \citep{van2000mean}, respectively. Their study showed that the different codes performed very differently, e.g., some of the codes reconstructed successfully the basic magnetic configurations in the reference model, while others failed. Although some of the tested codes could qualitatively match the reference solution, the quantitative differences were quite significant. In a following joint test, \citet{derosa2009critical} found that although many NLFFF codes perform well with test cases, their effectiveness diminishes when applied to actual solar data. Different NLFFF codes produce significantly different field line configurations and provide varying estimates of magnetic free energy within the coronal volume for the same solar active region. Therefore, it is still a hot research topic to advance the NLFFF codes for reliable and sophisticated extrapolations. For example, very recently the machine learning techniques such as the physics-informed neural network method are also applied to NLFFF codes \citep{jarolim2023probing, jarolim2024advancing}.  

Among the currently available methods, the magneto-frictional (MF) method \citep{chodura19813d}, a special simplified version of the MHD relaxation method, has gained popularity for a long time due to its computational efficiency. The basic assumption of the MF method is taking the plasma as a fictitious non-Newtonian fluid by replacing the viscous dissipation term in the momentum equation with a frictional force. Furthermore, aided with a zero $\beta$ simplification (i.e., neglecting the gas pressure and gravity) and omitting the inertial terms, the MHD momentum equation is reduced to 
\begin{align} 
\nu\mathbf{v}=\mathbf{J}\times\mathbf{B}, 
\label{eq:MF}
\end{align}
where $\nu$ is the frictional coefficient. It simply means that the Lorentz force is balanced by the frictional force in the relaxation process, for which one only needs to solve the magnetic induction equation $ \partial \mathbf{B}/\partial t =\nabla \times (\mathbf{v} \times \mathbf{B}) $ with the velocity explicitly given by $\mathbf{v}=\mathbf{J}\times\mathbf{B}/\nu$. Application of the MF model is quite common in solar physics; there have been many studies with modelling of either static independent force-free equilibria \citep{roumeliotis1996stress, valori2005extrapolation, valori2010testing, guo2016magneto} or quasi-static evolution of the coronal magnetic field \citep{van2000mean, mackay2012sun, mackay2011modeling, cheung2012method, pomoell2019time, hoeksema2020coronal, yardley2021determining, afanasyev2023hybrid}.

Nevertheless, the non-Newtonian fluid assumption in the MF model has serious issues that may render the results dubious, which was first pointed out by \citet{low2013newtonian}. For example, he showed that the MF model “gives incorrect solutions to a broad class of physically meaningful initial-boundary problems for magnetic fields with neutral points”. This is simply because that at the neutral (or null) points (i.e., locations where $\mathbf{B} = 0$) the MF velocity $\mathbf{v}=\mathbf{J}\times\mathbf{B}/\nu$ will always be zero and therefore the null points will never be advected, while generally the null points need to move during the relaxation. More generally, \citet{low2013newtonian} argues that the MF model requires “that the fluid can only move in the direction of the Lorentz force at every space point. In a physical fluid, a relaxation to a minimum-energy state of the magnetic field involves a richer set of fluid displacements, along as well as against the Lorentz force.” 

Recently, \citet{yeates2022limitations} illustrated specifically some critical limitations of the MF model by a simple one-dimensional experiment in which the initial magnetic field is given by $B_x=B_z=0, B_y=1/2+\sin{\pi x}$ and periodic boundary conditions are used. He showed that the MF model could lead to substantial breakdown of magnetic flux conservation in the calculation, even without resistivity. Theoretically, for an ideal relaxation under zero-$\beta$ assumption, the field will eventually reach a minimum energy state, in which the field strength is uniform and current sheets (i.e., tangential discontinuities) form at the interfaces of the antiparallel field lines. The total unsigned magnetic flux conserves in the relaxation. \citet{yeates2022limitations} simulated the relaxation process using the MF model with different choices of the frictional coefficient $\nu$. Using a constant $\nu$, the total unsigned magnetic flux shows an ever decrease before reaching an equilibrium. In the solution with $\nu$ proportional to $B^2$, a common choice by many MF codes for the aim of speeding up the relaxation in weak field regions, the total unsigned magnetic flux is lost by around 30\% and the magnetic energy is lost by 50\% as compared with the theoretical values. Such a loss of magnetic flux and energy is not due to numerical diffusion since they used a very high resolution and a low-diffusive scheme in which the diffusion is assured to be negligible. Even worse, the field topology of the solution is totally changed as no current sheet is present and all the field lines direct in the same direction. 

Regarding the aforementioned serious limitations of the MF model, and that null points and current sheets are prevalent in the coronal magnetic field, both \citet{low2013newtonian} and \citet{yeates2022limitations} advocate to use the viscous relaxation model, in which the classical Newtonian viscosity is used to specific the velocity by 
\begin{align} 
- \mu \nabla^2 \mathbf{v} = (\nabla \times \mathbf{B}) \times \mathbf{B}.
\label{eq:mvr}
\end{align}
For example, \citet{yeates2022limitations} concluded that the expected minimum energy state is obtained correctly with this viscous relaxation method. Furthermore, unlike the MF model, the viscous relaxation method evolves rapidly toward the correct solution. Actually, some previous NLFFF codes (e.g., \citealp{mikic1994deducing,jiang2011reconstruction,inoue2013nonlinear}) have been developed based on the viscous MHD relaxation model with different levels of simplification made to the standard viscous MHD equations. For example, \citet{jiang2011reconstruction} solved the full MHD equations, while \citet{mikic1994deducing} and \citet{inoue2013nonlinear} solved the zero-beta MHD equations by neglecting the gas pressure. Here Eq.~\eqref{eq:mvr} is also a zero-beta MHD momentum equation but further simplified by discarding the inertial term. Therefore it can be considered as the simplest form of the viscous MHD relaxation model.

Inspired by the aforementioned studies, here we present an implementation of the viscous relaxation method for the purpose of NLFFF extrapolations. Different from previous MHD relaxation codes, we employed a fully implicit scheme to solve the simplified MHD equations, which has the advantage of not being constrained by the Courant–Friedrichs–Lewy (CFL) condition. We note that although the implicit schemes were rarely used in the numerical models of NLFFF, some implicit and semi-implicit approaches have been used for modeling the global corona \citep{hu1984full,lionello1999stability}. Through finite difference discretization, we obtain a massive nonlinear system of equations, which is solved using the Portable, Extensible Toolkit for Scientific Computation (PETSc; \citealt{balay1997efficient}). Based on this approach, we developed a fully-implicit viscous-relaxation NLFFF (FIVR-NLFFF) code. Three representative tests are employed to validate the code, including the semi-analytical force-free field model of \citet{low1990modeling}, the magnetic flux rope model of \citet{titov1999basic}, and a highly-sheared force-free field containing a strong current sheet as formed in the MHD simulation of \citet{jiang2021fundamental}. The rest of this paper is as follows. In Section~\ref{sec:Methods}, we describe the viscous relaxation model and its numerical implementation. In Section~\ref{sec:Results}, we compare our extrapolation results with three reference NLFFF models. A brief conclusion is given in Section~\ref{sec:Conclusion}.

\section{Methods} 
\label{sec:Methods}
\subsection{The viscous relaxation model}
As mentioned in the Section~\ref{sec:Introduction}, the viscous relaxation model is derived from the full MHD equations by a set of assumptions, which are elaborated below. The viscous compressible MHD equations can be described as: 
\begin{align} 
\frac{\partial \rho}{\partial t} + \nabla \cdot (\rho \mathbf{v}) = 0 \label{eq:mass_conservation},\\
\rho \frac{D \mathbf{v}}{D t} = 
                            - \nabla p + \mathbf{J} \times \mathbf{B} 
                            + \mu_s \nabla^2 \mathbf{v} 
                            + \left( \frac{1}{3} \mu_s + \mu_b \right) 
                            \nabla ( \nabla \cdot \mathbf{v} ) \label{eq:momentum},\\ 
\frac{\partial \mathbf{B}}{\partial t} =
                            \nabla \times (\mathbf{v} \times \mathbf{B}) 
                            + \eta \nabla^2 \mathbf{B} \label{eq:magnetic_field}.
\end{align}
Here $\mu_s$, $\mu_b$, and $\eta$ represent the coefficients of shear and bulk viscosity and magnetic diffusivity, respectively. Generally, these equations must be supplemented with an internal energy equation along with a closure relation that relates internal energy and pressure. However, in regions where the plasma $\beta$ is very small (i.e., where thermal pressure is much lower than magnetic pressure), the pressure $p$ and its gradient $\nabla p$ can effectively be set to zero, and the internal energy equation can be ignored. Moreover, in the active regions where the NLFFF model can be applied, the corona plasma is extremely tenuous and strongly magnetized. Thus in the process of seeking an equilibrium solution, the inertial term $\rho D\mathbf{v}/Dt$ can be neglected and the continuity equation is also ignored \citep{bajer2013magnetic}.  

Based on the above assumptions, the governing equations for the viscous relaxation model are formulated as follows:
\begin{align}
 \frac{\partial \mathbf{B}}{\partial t } &=\nabla \times (\mathbf{v} \times \mathbf{B}) + \eta \nabla^2 \mathbf{B}, \label{eq:Governing_equation1} \\
 0&=(\nabla \times \mathbf{B}) \times \mathbf{B} + \mu \nabla^2 \mathbf{v} \label{eq:Governing_equation2} .
\end{align}
Eq.~\eqref{eq:Governing_equation1} governs the evolution of the magnetic field, accounting for both ideal advection and resistive diffusion. The diffusion term is retained here to smooth out small oscillations induced by numerical discretization and also to change the magnetic field topology by reconnection when needed. Eq.~\eqref{eq:Governing_equation2} represents a force-balance condition, where the Lorentz force is counteracted by the viscous term. We note that the shear and bulk viscosity in Eq.~\eqref{eq:momentum} are simplified from tensors to a scalar $\mu$ representing the viscosity coefficient. Compared to the MF method, we cannot directly derive the instantaneous velocity from the magnetic field and then substitute it into the magnetic induction equation. Therefore, we solve this coupled set of equations through a fully implicit scheme. 

\subsection{Numerical implementation}

Eq.~\eqref{eq:Governing_equation1} and Eq.~\eqref{eq:Governing_equation2} involve three types of vector operators: the cross product, the curl, and the Laplacian. Expanding them yields the following component forms:
\begin{align}
    \frac{\partial B_x}{\partial t } &= \frac{\partial(v_x B_y - v_y B_x)}{\partial y} - \frac{\partial (v_z B_x - v_x B_z)}{\partial z} + \eta ( \frac{\partial^2 B_x}{\partial x^2}+ \frac{\partial^2 B_x}{\partial y^2} + \frac{\partial^2 B_x}{\partial z^2}), \label{eq:sp1} \\ 
    \frac{\partial B_y}{\partial t } &= \frac{\partial(v_y B_z - v_z B_y)}{\partial z} - \frac{\partial (v_x B_z - v_z B_x)}{\partial x} + \eta ( \frac{\partial^2 B_y}{\partial x^2}+ \frac{\partial^2 B_y}{\partial y^2} + \frac{\partial^2 B_y}{\partial z^2}),\\
    \frac{\partial B_z}{\partial t } &= \frac{\partial(v_z B_x - v_x B_z)}{\partial x} - \frac{\partial (v_y B_x - v_x B_y)}{\partial y} + \eta ( \frac{\partial^2 B_z}{\partial x^2}+ \frac{\partial^2 B_z}{\partial y^2} + \frac{\partial^2 B_z}{\partial z^2}), \\
    0 &= (\frac{\partial B_x}{\partial z} - \frac{\partial B_z}{\partial x})B_z - (\frac{\partial B_y}{\partial x} - \frac{\partial B_x}{\partial y})B_y + \mu (\frac{\partial^2 v_x}{\partial x^2}+ \frac{\partial^2 v_x}{\partial y^2} + \frac{\partial^2 v_x}{\partial z^2}),\\
    0 &= (\frac{\partial B_y}{\partial x} - \frac{\partial B_x}{\partial y})B_x - (\frac{\partial B_z}{\partial y} - \frac{\partial B_y}{\partial z})B_z + \mu (\frac{\partial^2 v_y}{\partial x^2}+ \frac{\partial^2 v_y}{\partial y^2} + \frac{\partial^2 v_y}{\partial z^2}),\\
    0 &= (\frac{\partial B_z}{\partial y} - \frac{\partial B_y}{\partial z})B_y - (\frac{\partial B_x}{\partial z} - \frac{\partial B_z}{\partial x})B_x + \mu (\frac{\partial^2 v_z}{\partial x^2}+ \frac{\partial^2 v_z}{\partial y^2} + \frac{\partial^2 v_z}{\partial z^2}). \label{eq:sp6}
\end{align}
The equations are discretized on a 3D uniform grid. We employed the central finite difference method for spatial discretization due to its simplicity and ease of extension to higher-order accuracy. Additionally, the low numerical dissipation in this scheme makes it highly effective for capturing sharp discontinuities such as current sheets. For time discretization, we used the fully implicit backward Euler method (i.e., the right hand side of the above equations are assumed to be values at time level of $n+1$), which is unconditionally stable.

The expressions for the first and second order spatial derivatives using the second-order central difference scheme at the new time level $n+1$ are as follows:
\begin{align}
    \frac{\partial u}{\partial \xi} & \approx \frac{u_{i+1}^{n+1} - u_{i-1}^{n+1}}{2\Delta \xi},\\
    \frac{\partial^2 u}{\partial \xi^2} & \approx \frac{u_{i+1}^{n+1} - 2u_i^{n+1} + u_{i-1}^{n+1}}{\Delta \xi^2},
\end{align}
where $\xi$ represents one of the three spatial coordinates $(x,y,z)$, $\Delta \xi$ the numerical grid size, $u$ any one of the solution variables, and $i$ the grid index along this coordinate. The time derivative is approximated by the backward Euler scheme as:
\begin{align}
    \frac{\partial u}{\partial t} \approx \frac{u^{n+1} - u^n}{\Delta t},
\end{align}
where $\Delta t$ is the time step size. Therefore the equations are transformed into a large-scale nonlinear algebraic equation system for the unknowns $\mathbf{U}^{n+1}$:
\begin{align}
\frac{\mathbf{U}^{n+1} - \mathbf{U}^n}{\Delta t} = \mathbf{S} (\mathbf{U}^{n+1}) \label{eq:U_t_S},
\end{align}
where $\mathbf{U} = (B_x, B_y, B_z, v_x, v_y, v_z)$, and $\mathbf{S}$ represents the right hand side terms of Eqs.~\eqref{eq:sp1}--\eqref{eq:sp6}. 
This fully implicit approach requires solving a system of nonlinear algebraic equations at each time step, and the solution $\mathbf{U}$ is advanced step by step in time until it converges, i.e., $\mathbf{U}^{n+1} =\mathbf{U}^n$ and thus $\mathbf{S}(\mathbf{U}^{n+1}) = 0 $. However, by taking advantage of the unconditionally stable nature of the scheme, we can use a very large time step $\Delta t$ such that $(\mathbf{U}^{1}-\mathbf{U}^{0}) / \Delta t \approx 0$ and the entire problem is solved in a single time step. This means that we only need to solve a nonlinear algebraic system 
$(\mathbf{U} - \mathbf{U^0})/\Delta t = \mathbf{S(U)}$, or equivalently, $\mathbf{U}/\Delta t - \mathbf{S(U)} = \mathbf{U^0}/\Delta t$, where $\mathbf{U^0}$ is the initial guess and U is the final solution. This nonlinear system is solved by the PETSc using an iterative algorithm, which will be described in Section~\ref{2.3}.

An iterative algorithm often fails to converge when the initial guess is far from the exact solution; however, generating the initial guess from the corresponding linearized equation can greatly enhance both the robustness and efficiency of convergence \citep{casella2021choice}. In this paper, we used potential field as the initial guess following many previous NLFFF codes. Particularly, we use the Green's function method to compute the potential field based on the normal component $B_z$ specified at the bottom boundary. It corresponds to the lowest energy state of the magnetic field with the given bottom boundary $B_z$. We use the potential field throughout the entire domain as the initial condition, except at the bottom boundary where the vector magnetic field data, including the horizontal components, are applied.

The magnetic field and velocity adopt different boundary conditions. The magnetic field at the boundary surface is fixed, including its three vector components $B_x$, $B_y$, and $B_z$, with the bottom surface specified by the given vector magnetic field data and the other five surfaces set to the initial potential field values. For the boundary velocity on all six surfaces, we calculate it using Eq.~\eqref{eq:Governing_equation2}, which requires the first-order derivatives of the magnetic field and the second-order derivatives of the velocity. 

\begin{figure}
    \centering
    \centerline{\includegraphics[scale=0.5]{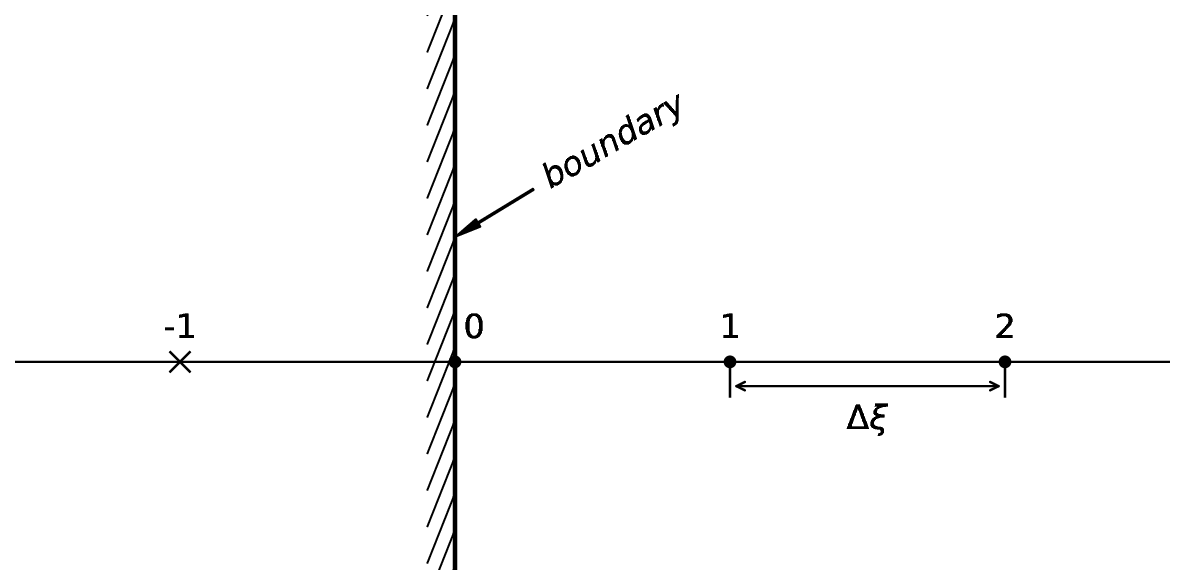}}
    \caption{The grid points near the left boundary surface.}
    \label{fig:1}
\end{figure}

Fig.~\ref{fig:1} shows the discrete points at and adjacent to the left boundary ($\xi = 0$). To ensure second-order accuracy, the first-order derivatives of the magnetic field is approximated using a  one-sided three-point scheme:
\begin{align}
\frac{\partial \mathbf{B}}{\partial \xi}\bigg|_{\xi = 0} \approx \frac{-3\mathbf{B}_0 + 4\mathbf{B}_{1} -\mathbf{B}_{2}}{2\Delta \xi}.
\end{align}
Additionally, we introduce a ghost point $\mathbf{v}_{-1}$ and enforce the condition that the normal derivative of the boundary velocity is zero, approximated by a second-order central difference:
\begin{align}
\frac{\partial \mathbf{v}}{\partial \xi}\bigg|_{\xi = 0} \approx \frac{\mathbf{v}_1 - \mathbf{v}_{-1}}{2\Delta \xi} = 0,
\end{align}
which leads to $\mathbf{v}_{-1} = \mathbf{v}_1$. With this relation, the second-order derivative of the velocity is approximated using a second-order central difference scheme:
\begin{align}
\frac{\partial^2 \mathbf{v}}{\partial \xi^2} \bigg|_{\xi = 0} \approx \frac{\mathbf{v}_1 - 2\mathbf{v}_0 + \mathbf{v}_{-1}}{\Delta \xi^2} = \frac{2\mathbf{v}_1 - 2\mathbf{v}_0}{\Delta \xi^2},
\end{align}
which allows the boundary velocity to satisfy the same governing equations as the interior nodes, without requiring points outside the computational domain.
The right boundary ($\xi = (N-1)\Delta \xi$) is treated in a similar fashion, where $N$ represents the number of nodes in $\xi$ direction, and the first-order derivatives of the magnetic field and the second-order derivatives of the velocity are as: 
\begin{align}
\frac{\partial \mathbf{B}}{\partial \xi}\bigg|_{\xi = (N-1)\Delta \xi} &\approx \frac{3\mathbf{B}_{N-1} - 4\mathbf{B}_{N-2} +\mathbf{B}_{N-3}}{2\Delta \xi},\\
\frac{\partial^2 \mathbf{v}}{\partial \xi^2} \bigg|_{\xi = (N-1)\Delta \xi} &\approx  \frac{2\mathbf{v}_{N-2} - 2\mathbf{v}_{N-1}}{\Delta \xi^2}.
\end{align}

Prior to the computation, the magnetic field is normalized by dividing by the maximum value of the initial field, ensuring consistent computational parameters across different cases. The viscosity coefficient is set to $\nu = 1$, and the magnetic diffusivity is typically chosen as $\eta = 0.005$. This choice effectively damps numerical oscillations without introducing significant dissipation. The computational domain uses a uniform grid with spacing $\Delta x = \Delta y = \Delta z = 1$. The entire problem is solved in just one time step with a large time step of $\Delta t = 12000$, which is sufficient for the magnetic field to reach a steady state. A even larger time step will result in unnecessary magnetic dissipation. In this study, all the test cases are computed using the same parameters.

\subsection{The PETSc solver}
\label{2.3}

To solve the large nonlinear system of Eq.~\eqref{eq:U_t_S}, we use PETSc, which provides a large suite of scalable parallel linear and nonlinear equation solvers, ODE integrators, and optimization algorithms for application codes written in C, C++, Fortran, and Python \citep{petsc-web-page}. In addition, PETSc includes support for managing parallel partial differential equation discretizations including parallel matrix and vector assembly routines. It greatly helps us for the code development process by avoiding the need to focus on implementation of specific algorithms or computational optimizations. Since we are solving for a steady-state solution in a regular cubic computational domain, a structured grid is used. In PETSc, the Domain Management Distributed Array (DMDA) is employed to store the discretized variable $\mathbf{U}$, which has a size of $ N_x \times N_y \times N_z \times 6 $. Here, $ N_x $, $ N_y $, and $ N_z $ are the number of grid points in the $x$, $y$, and $z$ directions, respectively, and 6 degrees of freedom represent the three components of the magnetic field and velocity \citep{balay1997efficient,may2016extreme}. The DMDA object automatically partitions the data across computational nodes and manages communication between them. Details regarding the implementation can be found in \cite{bueler2020petsc}.

To efficiently solve the asymmetric equations with extremely large sparse matrices, a hybrid approach combining the Jacobian-Free Newton-Krylov (JFNK) method with the Generalized Minimal Residual (GMRES) method was employed. The JFNK method, featuring a nested structure of Newton corrections and Krylov subspace iterations, eliminates the need for explicit Jacobian matrix computation by approximating it with small perturbations, thereby significantly accelerating the solution process \citep{chan1984nonlinearly,brown1990hybrid,knoll2004jacobian}. GMRES is chosen for its efficiency in minimizing residuals in indefinite and nonsymmetric linear systems, making it well-suited for the sparse matrix structures in this study \citep{saad1986gmres,saad1993flexible,zou2023gmres}. This approach ensures computational efficiency and scalability for the large-scale nonlinear system under investigation.

\section{Validation and results} 
\label{sec:Results}
To validate our code, we tested its performance using three reference models of force-free fields. These models include the Low and Lou's semi-analytic solution \citep{low1990modeling}, the Titov-Demoulin flux rope model \citep{titov1999basic}, and a strongly sheared arcade that contains a current sheet as modeled by \citet{jiang2021fundamental}.

\subsection{Figures of merit}
To quantitatively describe the degree of consistency between the model field and the extrapolated one, \citet{schrijver2006nonlinear} proposed five parameter metrics. Respectively, they are the vector correlation $C_{\rm vec}$
\begin{align}
    C_{\rm vec} = \frac{ \sum_i \mathbf{B}_i \cdot \mathbf{b}_i }{ \left( \sum_i |\mathbf{B}_i|^2 \sum_i |\mathbf{b}_i|^2 \right)^{1/2} },
\end{align}
the metric $C_{\text{CS}}$ based on the Cauchy-Schwarz inequality
\begin{align}
    C_{\rm CS} = \frac{1}{M} \sum_i \frac{\mathbf{B}_i \cdot \mathbf{b}_i}{\left | \mathbf{B}_i \right | \left | \mathbf{b}_i \right |},
\end{align}
the normalized and mean vector error metrics $E'_{\rm n}$ and $E'_{\rm m}$
\begin{align}
    E_{\rm n} = \frac{ \sum_i \left | \mathbf{b}_i - \mathbf{B}_i \right | }{ \sum_i \left | \mathbf{B}_i \right | }&; \quad E'_{\rm n} = 1 - E_{\rm n},\\
    E_{\rm m} = \frac{1}{M} \sum_i \frac{\left | \mathbf{b}_i - \mathbf{B}_i \right |}{\left | \mathbf{B}_i \right |}&; \quad E'_{\rm m} = 1 - E_{\rm m},
\end{align}
and the energy metric $\epsilon$
\begin{align}
    \epsilon  = \frac{\sum_i |\mathbf{b}_i|^2}{\sum_i |\mathbf{B}_i|^2},
\end{align}
where $\mathbf{B}_i$ and $\mathbf{b}_i$ represent the reference field and the extrapolated field, respectively. The variable $i$ stands for the indices of the grid points, and $M$ is the total number of grid points. It's evident that a precise extrapolation will result in all metrics equating to one in such definitions. The closer to unity the metrics are, the better the extrapolation is.

In addition to the above metrics, the two parameters $\mathrm{CWsin}$ and $\langle |f_i| \rangle$ are also used to evaluate the force-freeness and divergence-freeness of the extrapolation \citep{wheatland2000optimization}:
\begin{align}
    \mathrm{CWsin} &= \frac{\sum_i|\mathbf{J}_i|\sigma_i}{\sum_i|\mathbf{J}_i|}; \quad \sigma_i = \frac{|\mathbf{J}_i \times \mathbf{B}_i|}{|\mathbf{J}_i||\mathbf{B}_i|},\\
    \langle |f_i| \rangle &= \frac{1}{M} \sum_i \frac{(\nabla \cdot \mathbf{B})_i}{6|\mathbf{B}_i|/\Delta \xi}; \quad
    (\nabla\cdot\mathbf{B})_{i}=\sum_{\xi\in{x,y,z}}\frac{B_{\xi}^{\,i+e_\xi}-B_{\xi}^{\,i-e_\xi}}{2\,\Delta\xi},
\end{align}
where $e_{\xi}$ denotes the unit grid displacement along the $\xi$ direction. For a perfect force-free field and divergence-free field, these two parameters are both zero.

\subsection{Low and Lou's force-free model}
\label{sec:LL}
\citet{low1990modeling} proposed a closed-form magnetic field model which has been widely used as a reference for magnetic field extrapolation codes \citep{schrijver2006nonlinear}. The fields can be described in terms of solutions to a second-order, nonlinear ordinary differential equation:
\begin{align}
    (1-\cos^2{\theta} ) \frac{d^2P}{d(\cos{\theta} )^2}+n(n+1)P+a^2\frac{1+n}{n}P^{1+2/n}=0, \label{eq:LLP}
\end{align}
where $n$ and $a$ are constants. Then the magnetic field is given by
\begin{align}
\mathbf{B}=\frac{1}{r\sin{\theta}} \left (  \frac{1}{r}\frac{\partial A}{\partial\theta}\hat{r}-\frac{\partial A}{\partial r} \hat{\theta} + Q\hat{\phi} \right ),
\end{align}
where $A=P(\cos{\theta})/r^n$ and $Q=aA^{1+1/n}$. 
The solution $P$ of Eq.~\eqref{eq:LLP} is uniquely defined by eigenvalue $n$ and its corresponding number of nodes $m$. Although the configured magnetic field is expressed in spherical coordinates, a plane can be placed in space and rotated at an angle to take a boxed region above the plane, thus generating a unique test case for our code in Cartesian box. The position of the plane is characterized by two additional parameters, $l$ and $\phi$, respectively. Therefore, we can generate a test field using four parameters. Here we have set $n=1$, $m=1$, $l=0.3$, and $\phi=\pi/4$, which gives the same reference model used by many authors \citep{wheatland2000optimization, amari2006well, jiang2011reconstruction, jarolim2023probing}. The computational domain is $[-1, 1] \times [-1, 1] \times [0, 2]$, and the grid resolution is $64 \times 64 \times 64$. 

\begin{figure}
\centering
\centerline{\includegraphics[scale=0.6]{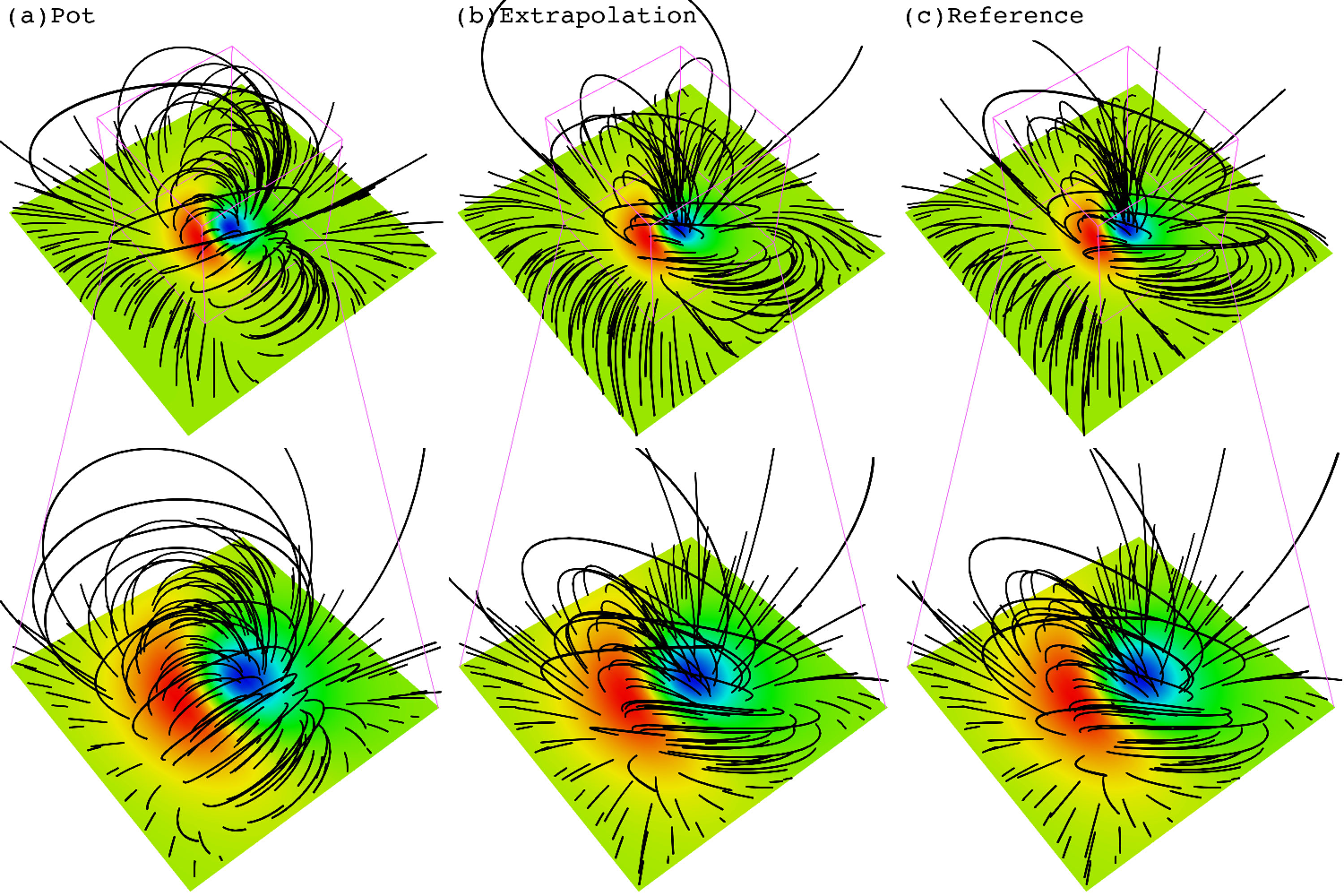}}
\caption{Magnetic field lines with pseudo-colored by value of $B_z$ (red for positive polarity and blue for negative polarity). They are the initial potential field (a), our extrapolation result (b), and Low and Lou's reference solution (c). The bottom row shows enlarged view of the central region in the top row.}
\label{fig:2}
\end{figure}

Fig.~\ref{fig:2} presents the potential field, the numerical extrapolation results, and the reference model solution. The plane displays the vertical magnetic field component $B_z$ at the bottom boundary, and the black curves indicate magnetic field lines anchored at the bottom boundary. The extrapolated magnetic field exhibits a sheared force-free configuration that is largely consistent with the reference solution, particularly in the central region (bottom row). Differences between the extrapolation and the reference model are primarily observed near the upper and side boundaries, which is attributable to the use of fixed potential field values as boundary conditions rather than the reference solution. 

\autoref{tbl:1} provides a quantitative comparison between the extrapolated field and the reference one in the central region. The four metrics of vector field correlation as well as the energy metric are very close to unity, indicating overall good agreement between the two fields. The degree of force-freeness of the extrapolated field is comparable to that of the reference field, and the magnetic field divergence remains at an acceptably low level.

\begin{table}
    \centering
    \caption{Quantitative comparison of the extrapolated magnetic field with the reference one for the Low and Lou's force-free model.}
    \begin{tabular}{lccccccc}     
        \hline
        Model & $C_{\mathrm{vec}}$ & $C_{\mathrm{CS}}$ & $E'_{\mathrm{n}}$ & $E'_{\mathrm{m}}$ & $\epsilon $ & $\rm CWsin$ & $\langle\vert f_i \vert \rangle \times 10^{-4}$ \\
        \hline
        Reference           & 1.00 & 1.00 & 1.00 & 1.00 & 1.00 & 0.04 & 2.77 \\
        Our Result          & 1.00 & 1.00 & 0.96 & 0.92 & 1.02 & 0.04 & 4.14 \\
        Potential Field     & 0.86 & 0.85 & 0.49 & 0.43 & 0.78 & 0.97 & 2.59 \\
        \hline
    \end{tabular}
    \label{tbl:1}
\end{table}

\begin{figure}
    \centerline{\includegraphics[scale=0.6]{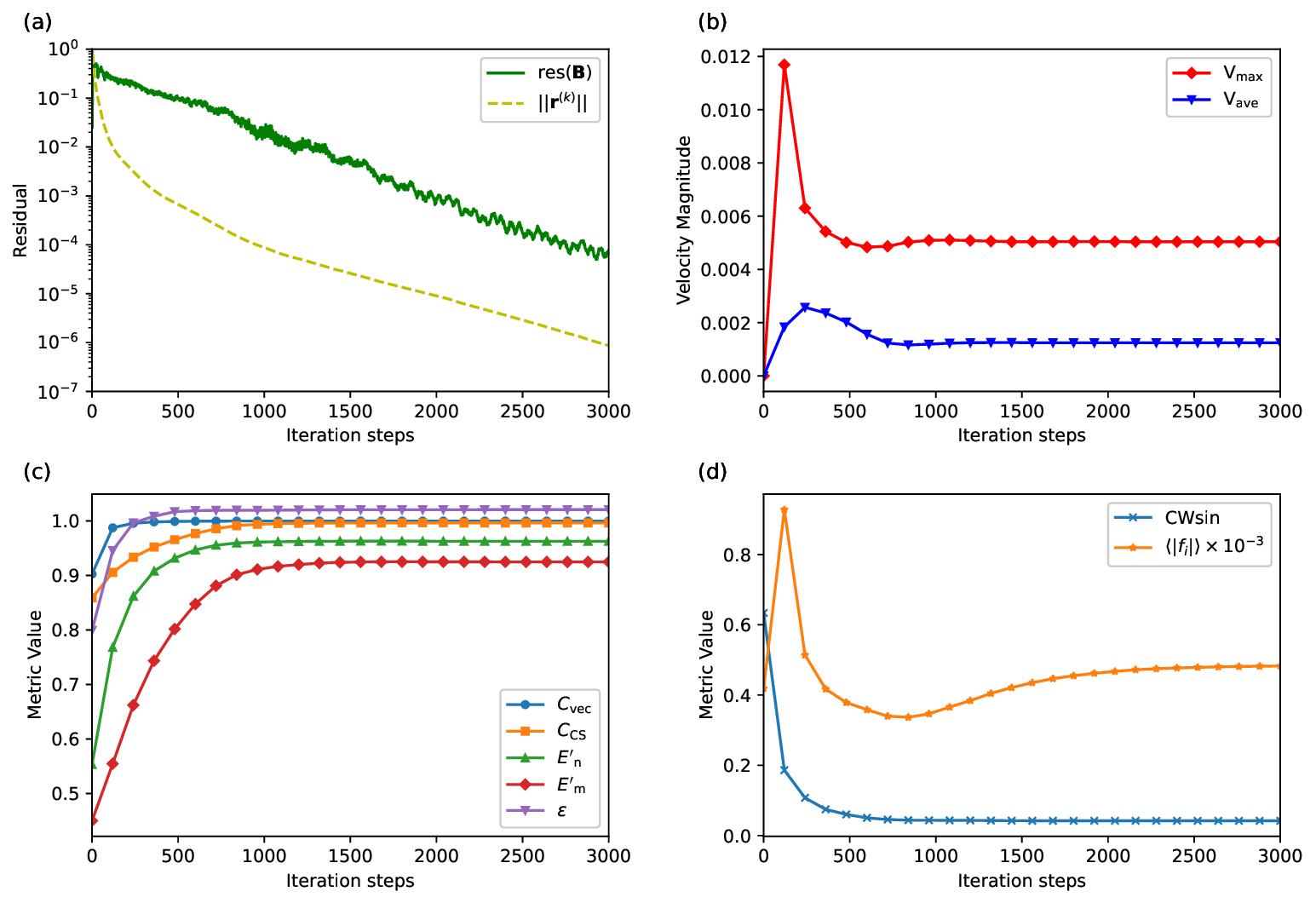}}
    \caption{History of the iterative solving process: (a) residual of the central magnetic field res$(\mathbf{B})$ and residual $||\mathbf{r}^{(k)}||$ of the governing equation; (b) maximum and average velocities; (c) and (d) figures of merit for the central region.}
    \label{fig:3}
\end{figure}

Fig.~\ref{fig:3} presents the evolution of key quantities during the iteration process in the PETSc solver. Both the residual of the central magnetic field,
\begin{align}
\mathrm{res}(\mathbf{B}) = \sqrt{\frac{1}{3}\sum_{i} \frac{|\mathbf{B}_{i}^{(k+1)} - \mathbf{B}_{i}^{(k)}|^2}{|\mathbf{B}_{i}^{(k+1)}|^2}},
\end{align}
and the residual norm $||\mathbf{r}^{(k)}|| = ||\mathbf{F}(\mathbf{U}^{(k)})||$ decrease steadily, where $k$ is the Newton iteration step and $\mathbf{F}$ denotes the nonlinear system. The maximum and average velocities approach low values, indicating relaxation toward a static state. All the metrics for correlation with the reference model increase monotonically and saturate within about 1000 steps, demonstrating a fast convergence to the target force-free field. Magnetic divergence initially increases slightly but remains low overall. Therefore, convergence is defined as when either $\mathrm{res}(\mathbf{B}) \leq 7 \times 10^{-3}$ or $||\mathbf{r}^{(k)}|| / ||\mathbf{r}^{(0)}|| \leq 10^{-4}$ is satisfied. We adopt the same convergence criterion for the remaining two tests.

\subsection{Titov-Démoulin magnetic flux rope model}

To simulate the twisted magnetic configurations in corona, \citet{titov1999basic} proposed a model with a force-free flux tube whose arc-like body is embedded into an external potential magnetic field. This model is commonly used for study of the structures of eruptive magnetic fields \citep{torok2005confined,kliem2006torus,liu2024model}. The configuration consists of a force-free circular tube with current $I$, a pair of magnetic charges $q$, $-q$ and a line current $I_0$ running along the tube's axis. By adjusting its parameters, the model can be adapted to depict various phases of a twisted flux tube's emergence from beneath the photosphere into the corona. There are two distinct configurations: one featuring a bald patch separatrix surface (BPSS) and the other incorporating a hyperbolic flux tube (HFT), which represent the stages of partial and complete emergence of the flux rope, respectively. See a detailed description of the model and its parameterization in \citep{titov1999basic,valori2010testing,titov2014method,jiang2016testing}.

Here we used the same test models from \citet{valori2010testing} who conducted a comprehensive evaluation of their extrapolation code with the TD model of four distinct stable configurations: the Low-HFT, High-HFT, No-HFT, and BP cases. In this study, we employ the same reference data, with a grid resolution of $102 \times 166 \times 75$. To minimize the influence of the outer boundaries on the computational results and to facilitate comparison, we excluded the numerical results near the top boundary and within the outermost 20 grid cells. This treatment is consistent with the approaches used by \citet{valori2010testing} and \citet{jiang2016testing}.

\begin{figure}
    \centerline{\includegraphics[scale=0.7]{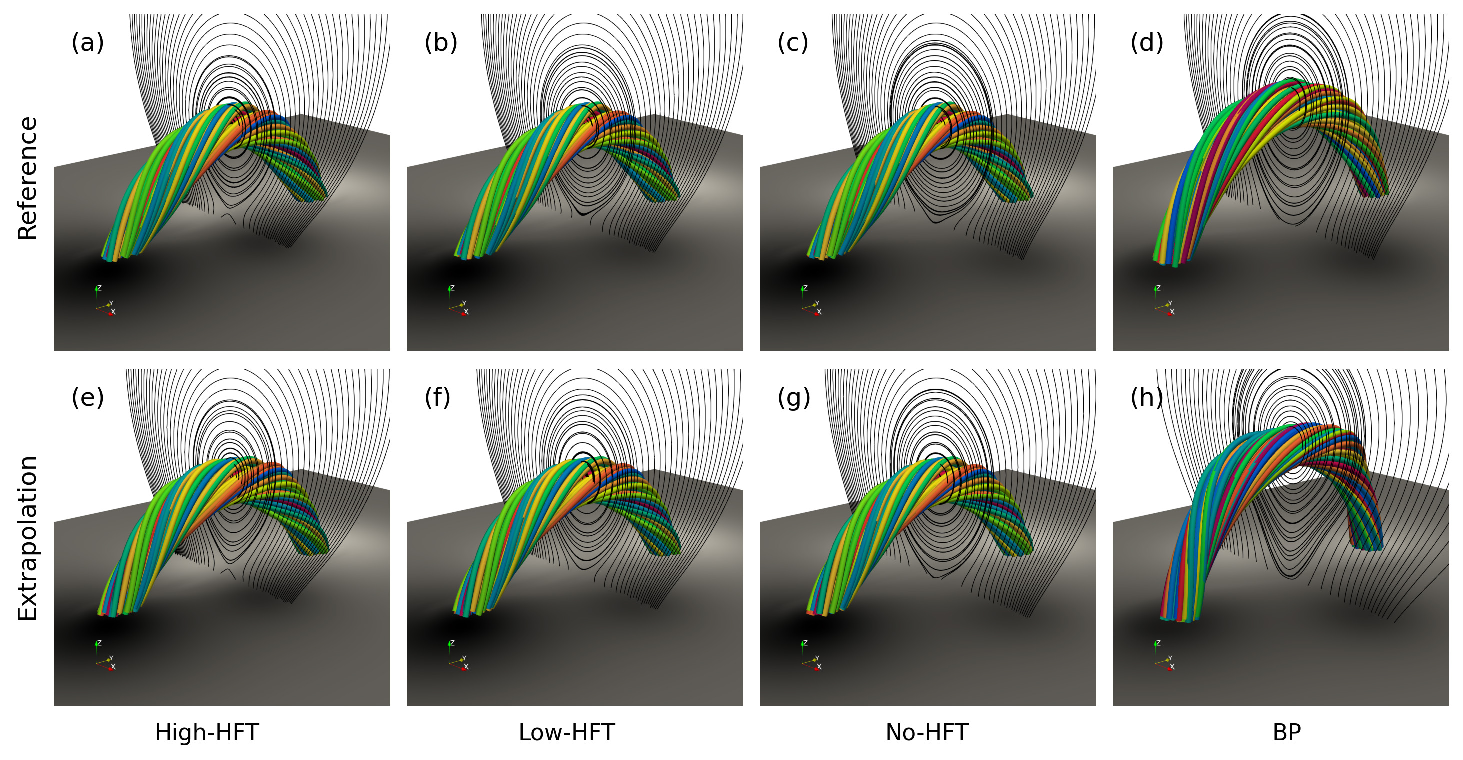}}
    \caption{Magnetic field configuration for the TD model test case. The colored lines represent magnetic field lines passing through a selected central spherical point. The bottom panel shows the distribution of $B_z$, while the black lines in the cutting plane indicate the 2D magnetic field lines in the $y = 0$ plane. Top row: reference magnetic field for (a) High-HFT, (b) Low-HFT, (c) No-HFT, and (d) BP. Bottom row: corresponding extrapolated results.}
    \label{fig:4}
\end{figure}

Fig.~\ref{fig:4} visually illustrates the configuration of the TD model. We selected a spherical scatter point at the center of the magnetic flux rope as the reference point and depicted the structure of the magnetic rope. By distinguishing the magnetic structures with different colors, we confirmed that our results (bottom) are nearly identical to the reference (top). The two-dimensional (2D) distribution of the magnetic field lines in the middle plane (black lines) highlights the locations of the topological structures, i.e., BP and HFT. The extrapolations reproduced almost the same locations for these structures as those of the reference model.

We used the same evaluation metrics as before to quantitatively compare our extrapolation results with the reference values. Furthermore, to compare our result with the previous published ones \citep{valori2010testing,jiang2016testing}, we compute the same parameters related to the magnetic topology as employed by them. These are the relative errors in the apex heights of the flux rope axis (FRA) and the HFT (if it exists) between the extrapolated and reference fields. Both apex heights can be roughly seen as the inversion points of $ B_x(0, 0, z) $ which is a good estimate of the poloidal field component along the line-symmetric $z$-axis. We also calculated the magnetic field energy error between the extrapolated result and the reference value. For the above indicators, the smaller the value is, the better the extrapolation is.

The results are given in \autoref{tbl:2}. Compared to the results of \citet{valori2010testing} and \citet{jiang2016testing}, our numerical method yields slightly higher values. However, the results remain within a rather small range. The BP configuration shows slightly larger differences from the reference compared to the other configurations. Fine-tuning of the resistivity or using a high resolution may yield better results.

\begin{table}
    \centering
    \caption{Extrapolation metrics results of the Titov–Démoulin magnetic flux rope model.}
    \label{tbl:2}
    \begin{tabular}{clcccccc}
    \hline
    Type&Field& $E_{\mathrm{m}}$&$\mathrm{CWsin}\times10^2$& $\langle\vert f_i \vert \rangle \times10^{-5}$ & HFT apex&FRA apex  &$E_{\mathrm{mag}}$\\ 
    \hline
    &Reference& 0& 1.442& 5.676& 0& 0&0\\ 
    High-HFT&Extrapolation&  0.046& 5.676& 4.459& 13.01\%& 4.87\%  &0.22\%\\ 
    &Valori& 0.044& 1.548& 6.044& 1.92\%& 4.20\%  &0.22\%\\
    &Jiang& 0.020& 2.07& 4.06& 5.18\%& 5.03\%&0.75\%\\ 
    \hline
    &Reference& 0& 1.274& 5.396& 0& 0&0\\ 
    Low-HFT&Extrapolation& 0.048& 5.681& 4.197& 16.51\%& 3.57\%  &0.30\%\\
    &Valori& 0.035& 0.733& 6.899& 0.63\%& 0.62\%  &0.37\%\\
    &Jiang& 0.015& 2.04& 3.28& 15.5\%& 1.38\%&0.98\%\\
    \hline
    &Reference& 0& 1.167& 5.351& $\cdots$& 0&0\\
    No-HFT&Extrapolation& 0.055& 5.635& 4.020& $\cdots$& 4.39\%  &0.30\%\\
    &Valori& 0.033& 1.176& 10.28& $\cdots$& 2.65\%  &0.75\%\\
    &Jiang& 0.016& 2.25& 3.52& $\cdots$& 0.05\%&1.1\%\\
    \hline
    &Reference& 0& 0.571& 7.759& $\cdots$& 0&0\\
    BP&Extrapolation& 0.183& 6.897& 10.687& $\cdots$& 10.15\%  &0.25\%\\
    &Valori& 0.054& 0.800& 10.05& $\cdots$& 0.29\%  &0.03\%\\ 
    &Jiang& 0.116& 5.41& 12.7& $\cdots$& 22.1\%&2.1\%\\
    \hline
    \end{tabular}
\end{table}

\subsection{A strongly sheared arcade containing a current sheet}
\begin{figure}
    \centerline{\includegraphics[scale=0.4]{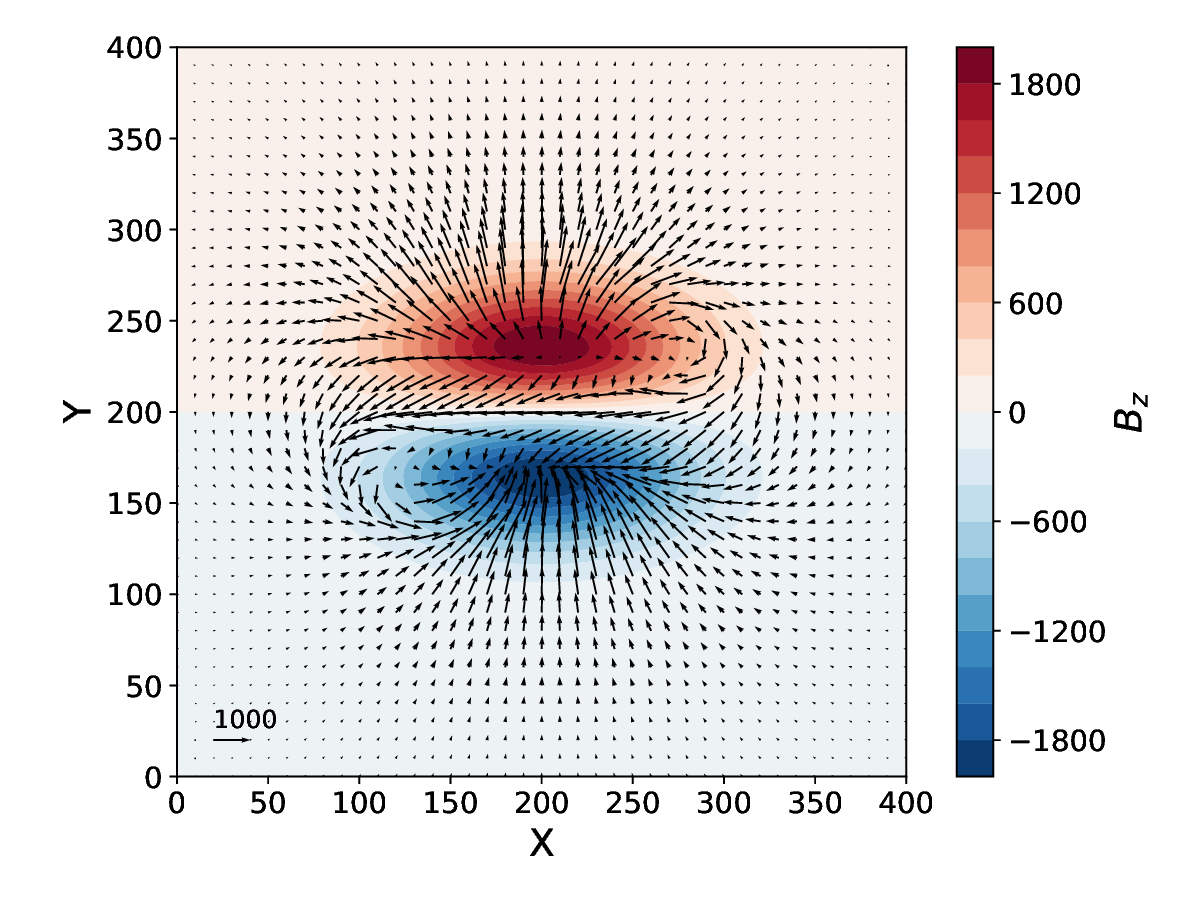}}
    \caption{The vector magnetic field data at the bottom boundary of the strongly sheared arcade model. The background shows the vertical component $B_z$, and the arrows show the horizontal component ($B_x, B_y$).}
    \label{fig:5}
\end{figure}

With a high accuracy MHD simulation, \citet{jiang2021fundamental} have demonstrated that the continual shearing of a magnetic arcade will result in a current sheet, and reconnection in this current sheet will produce an eruption. Therefore, the field immediately prior to the eruption in that simulation can be used to test whether a code can extrapolate such a pre-eruption field that contains a current sheet. Specifically we take the field at $t = 218~\mathrm{min}$ (3 min before the eruption onset) in \citet{jiang2021fundamental}'s simulation as our reference model. The original solution is obtained on an AMR grid with highest resolution of 90 km. We extract from the original solution a uniform grid of $401^3$ with resolution of 360 km. Using the vector magnetic field data of the bottom boundary which is shown in Fig.~\ref{fig:5} as the only input, we aim to extrapolate the coronal magnetic field containing the current sheet. It should be noted that extrapolating the field with a current sheet is challenging, because the thin current layer will broaden due to numerical dissipation, which induces numerical reconnection that prevents the formation of a structure with sufficiently large degree of magnetic shear. Therefore, a current sheet can only form in a quasi-static manner if the numerical diffusivity is sufficiently low. 

For this problem, we employed a multi-grid approach to speed up the computation. First, we performed the computation on a coarse grid ($101^3$), with the bottom boundary data coarsened by four times from the original high resolution grid. The resulted solution was then interpolated using cubic splines to an intermediate grid ($201^3$), on which we performed the second computation. Finally, the intermediate solution was interpolated onto the original fine grid ($401^3$), where the final relaxation was performed to obtain the converged result.

\begin{figure}
    \centerline{\includegraphics[scale=0.7]{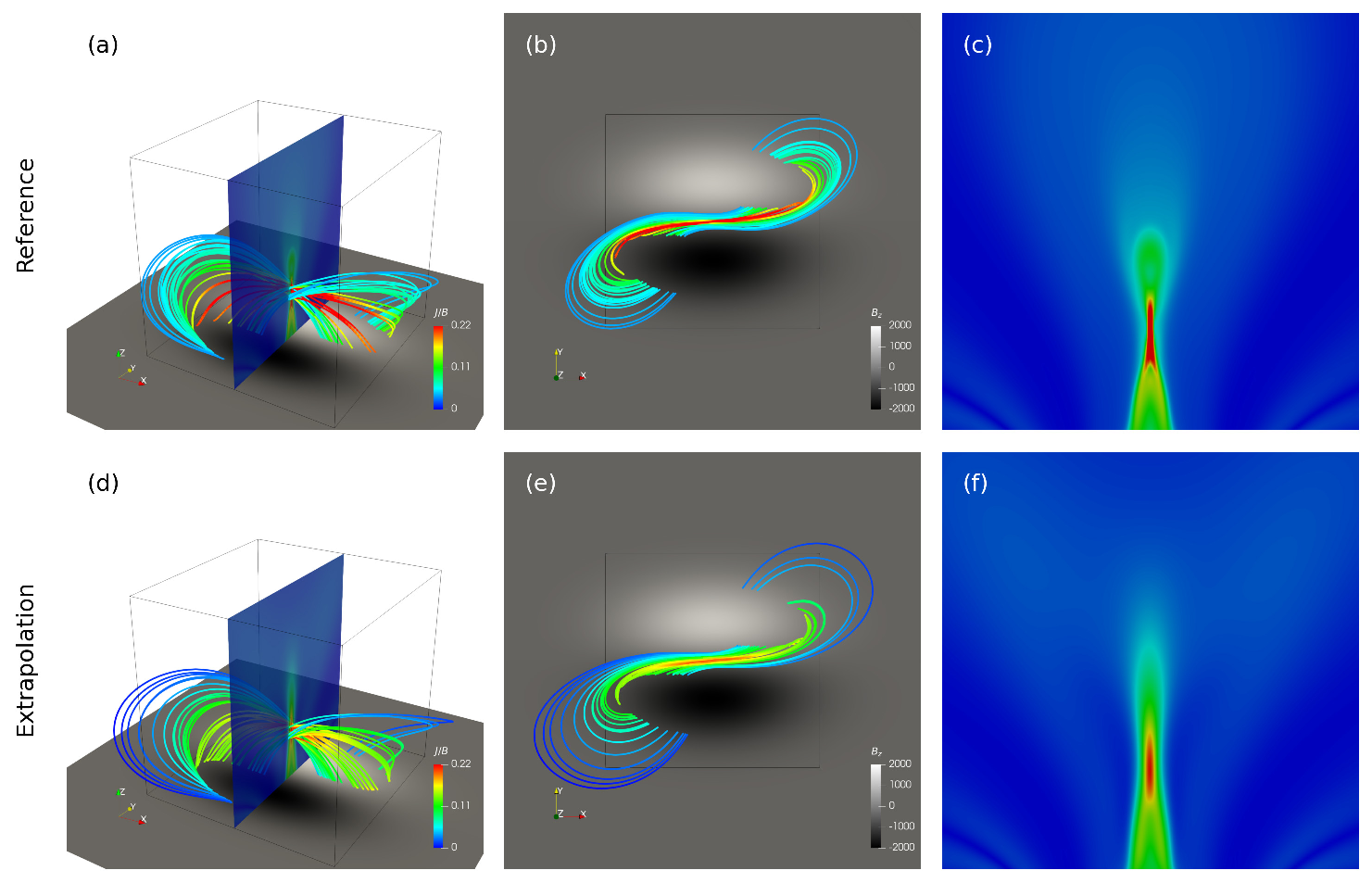}}
    \caption{Current density distribution of the strongly sheared arcade model. (a) 3D view of the reference model, with the bottom surface showing the distribution of $B_z$ on the $z=0$ plane, the vertical cross section showing the distribution of $J/B$, and lines indicating the magnetic field configuration; (b) top-down view of the reference model; (c) the distribution of $J/B$. (d), (e) and (f) are corresponding extrapolated results.}
    \label{fig:6}
\end{figure}

Fig.~\ref{fig:6} compares the magnetic field configuration of the extrapolated field with that of the reference model. A central vertical slice of $J/B$ is shown for the cross section of the current sheet. A symmetric seed points was selected around the center of the current sheet to trace the magnetic field lines, which reveal a reverse S-shaped structure and represent the highly sheared magnetic field configuration. Our extrapolation reproduced the current sheet structure in the core of the sheared field. Although the sheet is thicker than in reality due to grid resolution limits, the strong current region spans only 3 to 5 grid points. \autoref{tbl:3} shows the quantitative comparison based on the metrics. All the metrics are computed for the central area as denoted by the black box in Fig.~\ref{fig:6}. They show the extrapolated field reaches a highly force-free state, and the magnetic energy is almost identical to the reference one. 

\begin{table}
    \centering
    \caption{Quantitative comparison of the strongly sheared arcade reference in the central region.}
    \begin{tabular}{lccccccc}     
        \hline
        Model & $C_{\mathrm{vec}}$ & $C_{\mathrm{CS}}$ & $E'_{\mathrm{n}}$ & $E'_{\mathrm{m}}$ & $\epsilon $ & $\rm CWsin$ & $\langle\vert f_i \vert \rangle \times 10^{-4}$ \\
        \hline
        Reference           & 1.00 & 1.00 & 1.00 & 1.00 & 1.00 & 0.01 & 0.09 \\
        Extrapolation       & 0.99 & 0.97 & 0.89 & 0.79 & 0.98 & 0.02 & 0.34 \\
        Potential Field     & 0.68 & 0.70 & 0.27 & 0.22 & 0.56 & 0.99 & 0.02 \\
        \hline
    \end{tabular}
    \label{tbl:3}
\end{table}

\section{Conclusion}
\label{sec:Conclusion}
In this study, we have developed a FIVR-NLFFF code for extrapolation of the solar coronal magnetic field. It solves a simplified set of resistive and viscous MHD equation with gas pressure and inertial term neglected. It applies a fully implicit numerical scheme realized by a combination of the JFNK and GMRES methods. The PETSc toolkit is employed for the actual computation with efficiency and parallelization. The FIVR-NLFFF code demonstrates enhanced robustness in handling discontinuous features such as current sheets, which are critical structure in the initiation of solar eruptions. To our knowledge, none of the previously available NLFFF codes has considered the ability to extrapolate current sheets in the corona. In this regard, our code is unique.  Moreover, it offers a robust alternative to the commonly employed magneto-frictional method that could violate the conservation of magnetic flux and lead to incorrect magnetic topologies. Through systematic application to various reference models including the Low and Lou's force-free model, the TD magnetic flux rope model, and the strongly sheared arcade containing a current sheet, we have shown that the FIVR-NLFFF achieves high accuracy and numerical stability. Furthermore, since the strongly-sheared arcade model from \citet{jiang2021fundamental} were not strictly stable (i.e., only 3 minutes before eruption), the ability of our extrapolation to reproduce such an about-to-erupt state is appealing. For instance, an MHD simulation of the subsequent eruption can be readily performed by initializing it with the extrapolated field. These strengths position it as a promising tool for future investigations of solar magnetic phenomena, including flare initiation, coronal mass ejections, and other eruptive processes in the corona. Because the full implicit method in general guarantees the robustness (e.g., the strong numerical stability, and the positivity of the plasma quantities when plasma density and temperature are included), the implicit algorithm employed in this model could be used in future MHD simulation for the time-evolution of the active-region magnetic field. It is also worth noting that, although we have not explicitly addressed the issue of numerical magnetic divergence by using any divergence-cleaning approach or divergence-controlling term in the model equations, the magnetic divergence errors in all the test results are maintained within a reasonably small level. This is likely owing to the advantage of the full implicit scheme in time with central difference in space. We plan to incorporate divergence control terms as commonly adopted in other codes (e.g., \citealp{jiang2011reconstruction,inoue2013nonlinear}) in future work to further reduce the divergence errors.

While this paper has elaborated the basic methodology and validation of the code, its computational efficiency will be discussed in a following paper. We plan to optimize the computation by using preconditioned GMRES, and conduct a dedicated comparison of our method with other available NLFFF codes in terms of efficiency. Particularly, we will take the observed data as input to perform extrapolation of the realistic corona and compare the result with observations.

\section*{Acknowledgments}

This work is jointly supported by National Natural Science Foundation of China (NSFC 42174200), Shenzhen Science and Technology Program (Grant No. RCJC20210609104422048), Shenzhen Key Laboratory Launching Project (No.ZDSYS2021070\allowbreak2140800001), Guangdong Basic and Applied Basic Research Foundation (2023B1515040021).

% \nolinenumbers
%% Bibliography
%% Author year style
% \bibliographystyle{jasr-model5-names}
% \biboptions{authoryear}
% \bibliography{refs}

\end{document}